\documentclass[pra,twocolumn,nofootinbib]{revtex4-2}
\usepackage{amsmath}
\usepackage{amsfonts}
\usepackage{amssymb}
\usepackage{graphicx}%
\usepackage{color}

\begin{document}

\def\lbar{\lambda\hskip-4.5pt\vrule height4.6pt depth-4.3pt width4pt}

~~\leftline{CCU-WG-S/2022{\_02\_b}}
\\
\title{A dual concept of the angle in mathematics and practice}
\author{Savely~G.~Karshenboim}
\email{savely.karshenboim@mpq.mpg.de}
\affiliation{Ludwig-Maximilians-Universit{\"a}t, Fakult{\"a}t f\"ur Physik, 80799 M\"unchen, Germany}
\affiliation{Max-Planck-Institut f\"ur Quantenoptik, Garching, 85748, Germany}

\begin{abstract}
We consider the angle in mathematics and arrive at a conclusion that there are two concepts on the issue. One is a descriptive geometrical one, while the other is from functional analysis. They are somewhat different, allow for different options, and both are legitimate and in use. Their difference may cause certain confusions. While the `geometrical angle' allows for different choice of units, the `functional angle' is a purely dimensionless one, being related to the angle in radians. We consider possible options to resolve the problem as it concerns the units.
\end{abstract}

%\today

\maketitle

Recently a problem of the unit for the angle and related questions within the SI system \cite{SI} became a topic of a discussion in the metrological community (see, e.g., \cite{MP}). We believe that the problem is not purely terminological, but is in part a conceptual one.

\section{Two concepts of the angle}

In mathematics, physics, and more practical activity two somewhat different concepts of the angle are simultaneously in use. One is geometrical and the other is from functional analysis. Both concepts cover the angle and its functions, such as the sine and cosine. (Speaking about the angle one should also have in minds various related quantities (see Appendix~\ref{s:phase} for detail)).

\begin{figure}[thbp]
\begin{center}
\includegraphics[width=0.85\columnwidth]{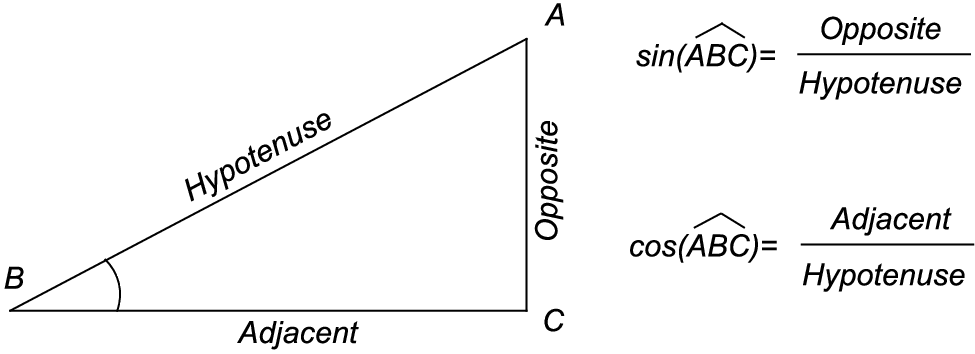}
\end{center}
\caption{The angle and its sine and cosine in a rectangular triangle. The angle itself is defined by a description of the triangle.
The functions of the angle are defined as a ratio of the length of the opposite or adjacent (to the angle) cathetus and the length of the hypotenuse.\\
As we see to describe an angle (of a triangle) in descriptive geometry we do not need any units. Neither they are required to find the key functions of the angle.
\label{fig:rect}}
\end{figure}

\underline{Concept \#1} considers the angle as a geometric object.
Once we have a triangle with the sides, the lengths of which are known, the angle is well defined. Such a definition does not require units [for the angle] immediately. The angle, defined so, directly allows for quantitative
evaluations. E.g., in a rectangular triangle with the known length of the sides one can find the sine and cosine of each angle still without any direct quantitative description of the angles themselves (see Fig.~\ref{fig:rect}). That is possible for any figures formed by segments of straight lines. (Note, an angle drawn on the floor, ground, or paper is a physical object that can be studied by means of physics. The compass and the straightedge can be also build as physical object. Using physical realization of geometrical objects and instrument we may in practice treat descriptive geometry as a part of physics.)

\begin{figure}[thbp]
\begin{center}
\includegraphics[width=0.45\columnwidth]{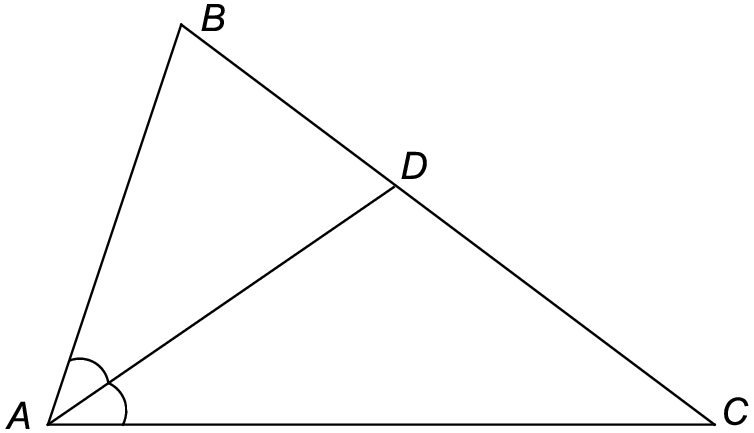}
\end{center}
\caption{Building a bisector and splitting an angle into two equal parts. That is a quantitative operation on an angle, that does not require any quantitative parameterization of it and may be performed with a compass and a straightedge, two standard master tools of descriptive geometry.
\label{fig:bisector}}
\end{figure}

The angle as a property of a figure allows in descriptive geometry for a number of operations and quantitative statements. Firstly, it is possible to sum and subtract angles, to build an angle which has the same value as a given one, to halve an angle (see Fig.~\ref{fig:bisector}) etc.

There are special objects, such as the full angle, the straight angle (a half of the full one), and the right angle (which is a quarter of the full one). There is a zero angle which is important for a description of mathematical relations between the angles. A possibility to make simple operations with the angles and the presence of the natural measures allow for a presentation of any angle exactly or approximately in terms of the full angle, as one of the possibilities of a quantitative description of angles.

Note, an angle drawn on the floor, ground, or paper is a physical object that can be studied by means of physics. The compass and the straightedge can be also build as physical object. Using physical realization of geometrical objects and instrument we may in practice treat descriptive geometry as a part of physics.

Rigorously speaking in metrological terms, probably we have to distinguish an object, the angle as its [quantitative] property, and a quantitative description of the property (or rather several of them).
All three are related, and we use term `angle' for all of them, but they are not the same.

\begin{figure}[thbp]
\begin{center}
\includegraphics[width=0.40\columnwidth]{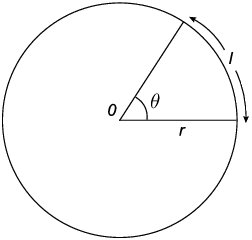}
\end{center}
\caption{The length of an arc is $l=\theta r$, where $\theta$ is a dimensionless parameterization of the [central] angle, such as the full angle is $2\pi$. Respectively, the length of the circumference, i.e., the arc related to the full angle, is $2\pi r$ and
number $\pi=3.14...$ can be calculated or measured, which makes the parameterization under question well defined.
\label{fig:arc}
}
\end{figure}

A construction, that consists of two lines, is an object (see, e.g., Fig.~\ref{fig:rect}). When we build an angle bisector we can say that we split the angle (as an object) in two angles with equal values (see Fig.~\ref{fig:bisector}), or alternatively we can say that we divide the angle (as a quantitative property) in two equal angles. We also deal with a quantitative property when calculate the sine and cosine for the angle of a triangle (see, e.g., Fig.~\ref{fig:rect}).
When we measure an angle in some units we consider a parameterization of the angle (see, e.g., Fig.~\ref{fig:arc}). We use the same term and in a sense equalize the concepts.

It is a complicated question what is the quantity, the quantitative property or its parameterization. (One should not confuse the choice of the units and the parameterization. An angle can be characterized by the related values of its sine and cosine. That is a parameterization, but it does not suggest any unit.---And that is a very efficient parameterization, that in particular allows for checking the equality of two angles and for their summation etc.\footnote{As a byproduct, the very possibility to adequately characterize the angle by its sine and cosine is a demonstration that there is no intrinsic dimensional unit hidden in the angles.}) The practice is that we often do not distinguish the property and its parameterization as two separate entities. As far as direct operations (such as splitting an angle in two by a bisector) on the quantitative properties are possible, one can consider the properties quantitatively, but still without any parameterization and separately from it.

That is not only the problem of the angle. In case of exclusive additive quantities we can often perform similar operations on quantitative properties in `natural terms' without specifying any parameterization, i.e., without a presentation of it as a product of a numerical value and a unit. One can double or halve the mass, weight (as a certain kind of force), length, volume etc. I should say that the mathematically well defined operations on the quantitative properties, without involving their numerical values, are preceding the introduction of the numerical operations. Rigorously speaking we often do not sum [numerical] values of two angles, masses, or volumes, we rather sum the angles, masses, or volumes themselves and interpret the result of the summation through the sum of the numerical values.

One should not confuse the parameterization of a quantity with its value. When we have two specific objects, say massive ones, we can speak about the values of their masses without any relation to the units and parameterizations. We can, e.g., state that the values of the mass of two different electrons are equal. The distance is a very general quantity, that is rather a geommetric one. We can parameterize it with a certain time period through the delay or return of echo of a certain signal. The distance between two specific points is a quantity value, that may be parameterized by different means or not parameterized at all. Note, that the presentation of a quantity value trough a numerical value and a unit definitely requires a certain parameterization, but the existence of neither the quantity nor the quantity does not.

In the meantime, the sine and cosine of the angle are defined as `pure' dimensionless quantities, values of which do not depend on how we parameterize (i.e., `measure') the angles. The sine and cosine can be considered as quantitative properties of the angle as an object (see Fig.~\ref{fig:rect}; the very notation $\sin(\widehat{ABC})$ is self explaining); they can also be considered as a certain non-numerical functions of the angle as a property of the object.

The `non-numerical' means that we can characterize the angle with no number, but considering it as an angle of a triangle or so, which is still sufficient for a determination of its sine and cosine (see, e.g., Fig.~\ref{fig:rect}).
Still the sine and cosine can also be considered as numerical functions of a certain parameterization of the angle, i.e., functions such that if we know their argument we can obtain their value through a certain
%algebraic
calculation without any geometric manipulations.

\begin{figure}[thbp]
\begin{center}
\includegraphics[width=0.50\columnwidth]{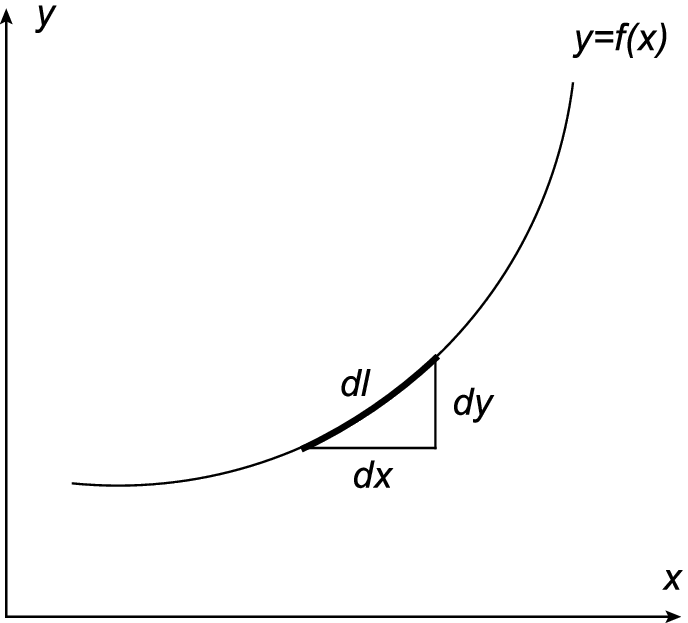}
\end{center}
\caption{
The length of an elementary segment of a curve described by $y=f(x)$ is $dl=\sqrt{dx^2+dy^2}$. The identity has sense only if $x$ and $y$ are quantities of the same kind.\\
To find the length of a finite segment one has to calculate integral $\int{\sqrt{1+(f'(x))^2}\,dx}$, which turns a geometric problem in a problem of functional analysis.
\label{fig:length}
}
\end{figure}

The geometric consideration is also {\em partly\/} valid for circles and their properties (cf. Fig~\ref{fig:arc}). However a rigorous description of curved lines and the related lengths and areas may at a certain stage involve limits, derivatives, and integrals, which makes this kind of geometric problems close to the functional approach (cf. Fig~\ref{fig:length}). Considering the length of a segment of a circle we can introduce the value of $l/r$ for an arc (see Fig~\ref{fig:arc}). It is referred to as the angle in radians. We may consider the usage of radians just as a comment that carries a kind of redundant piece of information used in order to avoid a possible confusion.

We may also consider word `radian' as the name of a specific angle as a geometric object, such as $l=r$. In this sense the radian similarly to the full angle is a natural measure of the angles that exists in descriptive geometry. Such a definition allows ones to introduce $\pi$ as a half length of a circumference with radius $r=1$ (as many mathematicians would say) or as the $l/r$ ratio related to the straight angle. It does {\em not\/} allow by itself to find the value of $\pi$, but it opens the door for subsequent measurements with elements of the descriptive geometry, realized in the physical world, or for subsequent calculations of various kinds.

Within the descriptive-geometry approach it is natural to compare various angles and to express their values in terms of natural angles, such as the full angle or the radian. In other words, it is natural to introduce certain reference values or units for the angle. Many metrological documents (see, e.g., \cite{SI}) consider a presentation of a quantity as a product of a numerical value and a reference. But none specifically tells that the `reference' is to be a dimensional one. The commonly used references are often used as a kind of units. We return to this problem later on in Sect.~\ref{s:ref}, but here we need to stress, that considering natural measures or references such as the full angle or the radian, we do not mean that the angle require a {\em dimensional\/} unit. Neither we mean by default that the full angle or the radian are dimensional.

\underline{Concept \#2} considers the angle as an object of the functional analysis. That means the presence of various relations between a numerical parameterization of the angle and numerical values of the functions of it. There are many relations of different kinds, that include derivatives, integrals, differential equations, Taylor series, such as
\begin{eqnarray}\label{eq:arccos}
\arccos(x)&=&\frac{\pi}2-\int_0^x{\frac{dx}{\sqrt{1-x^2}}}\nonumber\\
&=&\frac\pi2 -\sum_{k=0}^\infty \frac{(2k)!\,x^{2k+1}}{4^k(k!)^2(2k+1)}\;,
\end{eqnarray}
which allow ones to directly or indirectly define trigonometric functions (including the inverse ones) in terms of rational functions. In such a case with the absence of any geometrical meaning, one can speak about $\cos(2)$ or $\arccos(1/2)$ without any references to any geometric structure. The sine, cosine, and angle naturally appear as `pure' dimensionless numbers. The mentioned relations and functions appear also in physics as relations between the physical quantities or as a description of their properties etc. Once we deal with [numerical] relations between the quantities with objects, such as $x$ and $\arccos(x)$ in (\ref{eq:arccos}), we have to consider them as dimensionless quantities, not as numerical values of some dimensional quantities. (Metrological documents usually stress that there may be relations between the quantities or, alternatively, relations between their numerical values and separately between their units. But the relations on quantities should not contain the numerical values of the quantities by themselves, which is the very base of the {\em quantity calculus\/}. The same concerns the angle in relation $l=\theta r$ in the caption to Fig.~\ref{fig:arc}. That in particular means that the angle measured in the radians (i.e., the numerical value of the angle when the unit is the radian) can be a part of an equation on quantities only if the radian is equal to unity and therefore the angle in the radians is just the angle.

In the meantime, we can realize the geometrical axioms in terms of a certain coordinate space. We can `draw' lines and circles by writing related equations. In particular, we find that the angle understood in terms of the functional analysis is the same as the one equal to $l/r$ (see Fig~\ref{fig:arc}), as follows from a consideration of the length of a curve as the integral as shown in Fig~\ref{fig:length}. We can also find the numerical value of $\pi$ which requires a certain knowledge of the mathematical analysis.
We see that two concepts are in principle compatible, at least in the case of a certain parameterization of the [`geometrical'] angle, namely, in the radians.

Two mentioned concepts are actually not two concepts of the angle, but rather consequences of two concepts of geometry, a descriptive one and an analytic one. The former is based on naturally existing objects (or their analog models), while the latter speaks in terms of mathematical analysis and a numerical models. In the former case we first physically observe the objects and next describe and parameterize them, which gives us a certain liberty in choice of the parameterizations. In the latter case we first define the parameter space and next `build' [virtual] objects in it mathematically modeling the reality.

\section{Mathematics and dimensional description}

Above we spoke about geometry and angles which might make an impression that we meant our `actual' [position-space] geometry, the one we observe around through our everyday life. In the meantime, sometimes it is said that in mathematics they consider only dimensionless values. Both ideas play an important role in metrological discussions on the angle. And they are both not entirely correct.

\begin{figure}[thbp]
\begin{center}
\includegraphics[width=0.50\columnwidth]{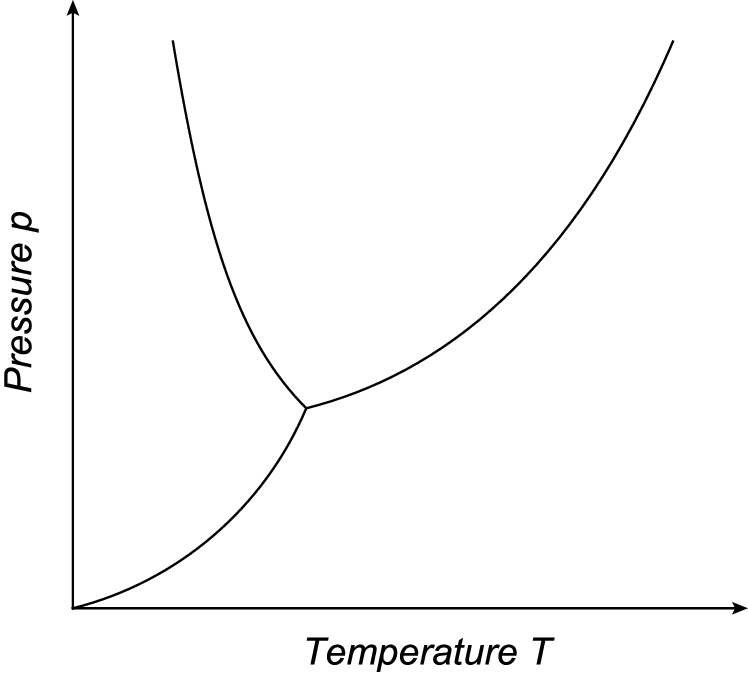}
\end{center}
\caption{The diagram that charts the phases of a substance.
The scales for the pressure $p$ and the temperature $T$ are unrelated and therefore the [apparent] angles in the plot have no physical meaning.\label{fig:triple}
}
\end{figure}

Many of us made plots. We may consider, e.g., a plot such as $p(T)$ (see, e.g., Fig.~\ref{fig:triple}). The plot does not completely satisfy the Euclidean geometry. That is because $p$ and $T$ are not compatible, since they are not `just' numbers. The set of possible values of $p$ is such as it is equivalent to nonnegative real numbers $\mathbb{R}_+$.
It is equivalent in a sense that we can build an one-to-one correspondence $p\to f(p)$, such as $f(p)$ is just a number and the correspondence maintains all the relations (equalities, inequalities) and linear operations.
The problem however is that we can build an infinite number of functions $f(p)$ which satisfy those conditions. Actually, a choice of $f(p)$ is a kind of a choice of the parameterization for $p$. In the case of the class of homogeneous linear functions the question about the parameterizations is reduced to the question of the units for $p$. (Considering $T$ similarly and introducing the linear functions that are not homogeneous additionally to the question of units (say, the degrees of Celsius or Fahrenheit) we also deal with choice of zero of $g(T)$ that differs from a parameterization to a parameterization.)

When we consider the $p-T$ space we note that set of possible $p$ is equivalent to $\mathbb{R}_+$ as well as set of possible $T$. But these two $\mathbb{R}_+$ sets are not comparable. There is no operation like $p+T$. That makes impossible to meaningfully consider angles. To consider them we have to build $f(p)$ and $g(T)$ which have values in the same set and therefore allows for an operation like $f(p)+g(T)$. Eventually that would allow for a consideration of angles. However, since we can make the correspondences $f(p)$ and $g(T)$ in many different ways the result is useless.

The same problem is with circles. We cannot draw a curve the points of which are at the same distance from a certain point, since the distances in $p$ and $T$ directions are not compatible. The noncompatibility of $p$ and $T$ does not preserve from linear operations in the $p-T$ parameter space, which makes the set of possible $(p,T)$ pairs to be a part of a vector space, but preserves it from the introduction of the scalar product, which is required for an Euclidian space.

It is often said that a vector is characterized by a magnitude and direction. We can separate those two characteristics when we describe a vector either in a kind of polar or spherical coordinates or use the magnitude and the so-called unity vector. In the $p-T$ space, we can introduce neither a meaningful magnitude nor a meaningful angles or a unity vector.

Considering a possibility of an operation like $p+T$ (or $f(p)+g(T)$ as an unambiguous one)
is a rigorous way how in mathematics they recognize that two quantities are of the same kind or not, which is the base property of the quantities that eventually delivers us the dimensions and units.

We may consider an $2d$ space with two `coordinates' of the same kind. Often that allows not only for linear operations but also for a scalar product which involves $a_1b_1 + a_2b_2$ where $a_i$ are coordinates of $\vec{a}$ and $b_i$ of $\vec{b}$. E.g., we can consider the momentum space where the coordinates are the components of the momentum. We understand that all the geometric statements are equally correct for the position space and the momentum space. That actually means that it is unimportant what is the dimension of the length of the line in Figs.~\ref{fig:rect}, \ref{fig:length}, and \ref{fig:arc}. It may be the one of the length (and the unit may be the meter), the one of the momentum (and the unit may be kg\,m/s), or anything else. It is important to recognize that the length of all the objects is of one dimension and it is {\em not\/} just a number. We can write
\[
{\rm `just~a~number'}+1
\]
and cannot write
\[
{\rm length}+1\;.
\]
It is also important to introduce the concept of the area and to recognize that the values of the area are of a dimension that is neither one of the length nor of `just' numbers, but we may build a meaningful function $h(l_1,l_2)$ the values of which would have a dimension of the area and proportional to $l_1$ and $l_2$. (Here we speak in algebraic terms about the area of a rectangle with sides $l_1,l_2$.)

In this sense the expressions of the functional approach to the angle and trigonometric functions such as (\ref{eq:arccos}) deal with `just' numbers and  $\theta=l/r$ from identity on the length of the arc (see Fig.~\ref{fig:arc}) is `just a number' as well. Expression\footnote{%
In mathematics we can write expressions like
\begin{eqnarray}
&&{\rm length}+{\rm length}\;,\nonumber\\
&&\frac{\rm length}{\rm length}+{\rm number}\;,\nonumber\\
&&{\rm length}\cdot{\rm length}+{\rm area}\;,
\end{eqnarray}
but cannot write
\begin{eqnarray}
&&{\rm length}+{\rm number}\;,\nonumber\\
&&{\rm length}+{\rm area}\;,
\end{eqnarray}
which set relations between the set of the numbers,
the set of values of the length, and the set of values of the area. More rigorously, we have to speak not about sets by themselves but about sets with several operations introduced, satisfying certain conditions.\\
The addition is introduced for all of these sets as well as the multiplication by a real number, which makes the one-dimensional vector space.
The possibility or impossibility for such an operation is equivalent to the notion of the dimension in physics and metrology.

In metrology the {\em quantity calculus\/} is a construction which recognizes different dimensions. The identities also assume a relation between the sets for different quantities, say, between the length and the area if the multiplication is involved.

While the consideration of the dimensions in mathematics is ignored, the {\em dimension calculus\/} is nevertheless present in an indirect way. A clear clarification whether the mathematical quantity is in fact dimensional is the operation on the quantity of interest
\[
{\rm quantity}+({\rm quantity})^2\;,
\]
which is allowed for a dimensionless quantity and not allowed for a dimensional one, which in mathematics is often understood but seldom spoken out. The Taylor expansion like the one in (\ref{eq:arccos}) indicates that the angle in functional analysis is dimensionless.

One may compare that with a notion of atomic units, that are dimensional and well-defined one, however, since each quantities have their well-recognized units, the notation in use is just `a.u.' for {\em all\/} atomic units, suggesting that the readers should themselves recognize the dimension of the quantity under question and use the related atomic unit.

Similarly, in many cases instead of dimensional values codes in practice contain only numerical values assuming that if everything is done properly the related identity on the units is fulfilled automatically. Technically however such an information is not spelled out inside the codes.
}
\[
{\rm angle}+1
\]
would have sense (cf. (\ref{eq:arccos})), which makes the [dimensionless] angle different from the [dimensional] length and area without any explicit acknowledgment of their dimensions.

Actually, the presence of quantities of unspecified dimensions does not belong exclusively to mathematics. In Lagrangian and Hamiltonian mechanics the generalized coordinates may be of any dimension. The same is with coordinates in general relativity. That happens often when we try to derive something in general. What is important it that a formalism (be it physical or mathematical) often does not specify dimensions of involved variables, but in any particular practical case of an application of the formalism the dimensions are specified in one or other way.

It actually is a great advantage that we can formulate some general solutions of certain kinds of problems. It is good that we can consider the harmonic oscillator with an amplitude of an unspecified dimension, instead of subsequently solving equations for a position oscillation, for an angle one, and additionally for an electric charge in a contour with a capacitor and an inductor.

We also enjoy such a possibility in metrology. In the data evaluation the value of $\chi^2$ is dimensionless, while a measured value of a datum (and its uncertainty) may be of any dimension. The normal distribution is of the same shape for any dimensional data as well as the rules of the propagation of the uncertainty etc.

It is worth also to mention that in metrology a quantity is defined (see, e.g., \cite{VIM}) as an one that can be presented as a product of a numerical value and a unit. That is possible if we are capable to make comparisons of quantities of the same kind and to present the results for their ratios, i.e., to perform certain operations without introducing any units. Similarly, as we see in the case of a bisector (see Fig.~\ref{fig:bisector}) we may perform certain operations on quantities without introducing any parameterization. (In the both cases we mean operation in our physical world on the quantities presented in natural terms, i.e., as properties of certain real objects.) The very possibility to introduce the numerical values and units is a direct consequence of the presence of such operations. Technically, if our intention is a quantitative description of objects and phenomena a `direct' parameterization of each property is not a necessity, but a possibility. E.g., for an angle an indirect parameterization, which completely describes the angle, is possible in many ways such as giving the values of the related sine and cosine (see, e.g., Fig.~\ref{fig:rect}) or the length of the related arc at a certain (possibly prearranged) value of the radius (see, e.g., Fig.~\ref{fig:arc}). The question of the units and dimensions appears when we discuss a kind of a direct parameterization and is in part avoidable.

It would be incorrect to say that in mathematics they used something else, less explicit, instead of the dimensions. One has to clearly understand what is the cause and what is the consequence. Roughly speaking, developing a quantitative description of natural phenomena one deals with individual values of quantitative properties which are members of certain sets with various relations, operations, and functions introduced on them. There are many different sets, different in a sense that they are not compatible, but distinguishable. One can consider the very introduction of the dimensions and units as tools. The introduction of the dimensions is one of possibilities to mark different sets and their compatibility, while the units is a way to correspond various sets to a set of numerical values, which are `just' numbers etc., i.e., to parameterize the sets.

\section{Reference values and dimensionless quantities\label{s:ref}}

A result of a measurement of a quantity is often represented as a product of a certain numerical value and a reference quantity of the same kind. We hesitate to use here term `unit', which is a core metrological term and there are a number of formal definitions of it, while the term `reference quantity' or `reference value' sufficiently describes something what actually is in use in practice and is more general.
When we say that one bought two dozens of eggs, the dozen serves as a reference value for counting the eggs. When we say about 12\% of the population of a town, the percent [of the population] serves as a reference value. There are many other examples when a certain problem has a natural or conventional reference value.

Dimensionless reference values play the same role as dimensional ones, except we often hesitate to call them `units'. However, if we do call them so, we start to speculate whether the related quantities are really dimensionless and whether those units are just numbers or something more.

It may be a long discussion whether the angle is a dimensional property, but in this section we would like to only demonstrate that measuring a certain quantity in terms of a certain reference value (does not matter whether we call it a unit) does not mean by itself that the quantity is dimensional. We also see that one can choose diverse reference values for dimensionless quantities, such as dozens or tens, percents or ppm.

\section{Special names of unity and the quantity calculus}

{\em Quantity calculus\/} is a special name of a procedure dealing with dimensional quantities and keeping their dimensions intact. The procedure is rigorously defined or at least is intended to be.

Meanwhile, there are various ideas of using a special name for the dimensionless quantities, i.e., a special name for unity. Introducing such a name may have advantages, e.g, for using prefixes for multiples and submultiples or to avoid a confusions in certain applications.

Using such a unit for data by themselves is possibly advantageous, but there is a problem in application of the {\em quantity calculus\/} on such data. We may introduce a certain unit which is equal to unity. That means that any procedure, which arbitrary introduces or drops it, is mathematically legitimate. Thinking about machine-reading procedures we arrive at a point where a machine should choose which of {\em several\/} legitimate operations to perform.

Most of [human-related] `ambiguities' come in practice through confusing different quantities, say, the frequency and the angular frequency, or the flux and the intensity. Sometimes those quantities have similar name, or very long names, short versions of which are very similar or even the same in a scientific `spoken' language. To distinguish different quantities by using the long names for them should not be a problem for a machine.

Another problem would appear if we deny the radian to be equal to unity, which relates to a consideration of the angle as a dimensional quantity. In such a case in some applications we should deal with the angle (as a dimensional quantity) measuring it in the radians, but in some others (see appendix~\ref{s:list}) we are to use the {\em numerical value\/} of the angle in the radians in equations with quantities. That undermines the rules of the quantity calculus.

\section{The angle: what are the options ?}

Prior of making any decision on the angle and the radian we have to realize what are the options. From the point of view of terminology the choice we have is the following.

Both concepts of the angle are present in practice and the related values are in massive use. We use geometrically understood angle when, e.g., we do a triangulation but we use the functional angle when, e.g., we consider a Taylor series of the cosine. (A list of examples of use of the dimensional angle in various areas of mathematics and physics is given in Appendix~\ref{s:list}, where we consider either equations related to the high-school level or University basics or to the base equations in somewhat advanced areas like quantum mechanics.)

Speaking about the functional approach, we can calculate angles and sines with help of integrals, Taylor series, or $l/r$. I.e., we can calculate the value of the angle from the functional concept. We can neither deny existence of the related values nor forbid their use. We can modify all the functional expressions to make them valid for a dimensional angle, e.g., for the angles in degrees. That is not a problem. The problem is that the value $l/r$ exists by itself, it is legitimate, it is in use, and it is called `angle'. The same is about the results of evaluations of (\ref{eq:arccos}) and many similar expansions. We may forbid to use term `angle' for them, but we cannot avoid the use of the related expressions. If the dimensionless angle from those expressions will not be named `angle' by the metrological community it will exist as the `customers angle'. The introduction of quantities is rather what the metrological community prefer to avoid. Their focus is the units (see, e.g., \cite{SI}) and names for the quantities in use, but not on their contents and definitions.

We also remind that the angle, considered in geometric terms is usually applied in pure geometric problems. There are many other applications, such as the phase angle in time series, the phase angle in applications which use complex numbers, the angle in other spaces than our $3d$ one. There are mathematical engines and tables and some of them use only angle of the functional concept. The dimensionless values introduced along the functional definitions are unavoidable and, being dimensionless, they {\em are\/} consistent with the SI\footnote{We have to distinguish the use of traditional approaches and units which is hard to stop for `social' reasons and the use of legitimate well-defined mathematical objects such $l/r$, which provides us with simplicity of notation. We can regulate the use of units, such a degrees or radians, but we cannot regulate the use of legitimately defined quantities except of constraining their names.}.

So, what can we do?

\underline{Option \#1.} We may introduce a clear terminological differences between two kinds of the angles, with, e.g., the `geometric angle' being a kind of a dimensional quantity (and we may discuss its units etc.) and the `functional angle' being just a number, which is equal to the numerical value of the geometric angle in the
radians. Both are to be allowed for use, but it may be recommended to use rather one of them than the other for certain occasions. Note, this solution suggests not only two angle quantities, but also a duplication of trigonometric functions as we discussed above (as non-numerical functions of a geometric object and as purely numerical functions technically (from a point of view of their definition and key relations) unrelated to any geometric object). That is the most clear once we consider the inverse trigonometric functions which would produce the [dimensional] geometric angles from `pure' numbers (cf. (\ref{eq:arccos})).

\underline{Option \#2.} The other option is to make two concepts explicitly consistent one to the other without any possible ambiguity. The latter may be reached only under the following conditions (that mostly follow from the functional approach that does not allow us too much room for flexibility).
\begin{itemize}
\item The angle is a dimensionless quantity, i.e., just a number.
\item The angle may be presented in terms of unit `radian', which satisfies the condition
$$1\;{\rm rad}=1\;,$$
i.e., the radian is just a special name of unity, used for some purposes (which is to be a subject of a certain regulation).
\item The other units for the angle should be considered as non-SI units being defined as certain multiples or submultiples of the radian, such as:
\begin{eqnarray}
{\rm full~angle}&=& 2\pi\;{\rm rad}\;,\nonumber\\
1^\circ&=&\frac{\pi\,{\rm rad}}{180}\;,\nonumber
\end{eqnarray}
etc.

Definition as a non-decimal multiple or submultiple is a common practice for non-SI units (such as the foot, the calorie, mm Hg) (cf. \cite{SI}).

There is a certain specifics in case of the dimensionless quantities. We do not call `\%' or `ppm' units, but sometimes treat them like units, despite they are in a sense just names of certain numbers. They are definitely in use as commonly accepted `reference values'.

(Technically, the degree is a 1/360th portion of the full angle, which is in its turn equal to $2\pi$ [radians]. So, it {\em is\/} a special number.)
\item Since we set $1\,{\rm rad}=1$ we can recommend on some occasions to keep `radian', but dropping it out should be also a valid option. Roughly speaking, we may limit the format of input and output data for certain applications but doing any calculation we should follow the {\em quantity calculus\/} which technically allows for dropping or re-introducing the radian at any stage. After all, if we intend to set a machine-readable data base we cannot forbid to perform a valid mathematical operation on its data.
\end{itemize}

Returning to the first option, we have to mention, that the geometric angle is a kind of a dimensional quantity. So, we need either to introduce it as a base or derived quantity of our system of units and quantities. If it is a derived one, say, as a ratio of two lengths, that would eventually make the angle a dimensionless quantity. Therefore, the choice is that the angle may be either a new base quantity or a dimensionless quantity. Dimensionless quantities are just the numbers. We indeed may use various reference values to present the results of their measurements. Those reference values, does not matter whether we call them units or not, should be the names of special numbers. E.g., we could consider $\pi$ as a unit, or a dozen as a such, or \%, etc. If the units are the name of numbers, which for the angle are actually well defined and exactly known ones, then we can always redefine such units to make the new dimensionless unit to be a name of unity, and once a dimensionless unit is unity it may be dropped out. The latter is a general statement on the units of the dimensionless quantities. Once we follow the {\em quantity calculus\/} there could be no independent meaningful dimensionless unit. They all can be expressed in the term of unity.

What is often considered as a dimensionless unit is rather a kind of a comment. Ten apples are not the same as ten oranges. But they have the same {\em number of fruits\/}. If the equation contains the number of fruits, one ten is equal to the other ten. (E.g., if we intend to provide {\em a\/} fruit per {\em a\/} lunch box.) If the equation, say, being a recipe for a certain fruit meal, specifically contains the number of apples or the number of oranges they are different. Many physical equations contain sums over the objects. If some objects are identical the equations contain a summation over the kinds of the objects. The number of identical objects of the same kind remains just a pure number. The difference is not in their numerical values, but in the context which reminds us that we may refer to different quantities. That is not related exclusively to dimensionless quantities. A kilogram of apples produces the same gravity as a kilogram of anything else. But a dessert recipe may specifically require a kilogram of apples. Speaking about pure scientific applications, one can also compare a physical and chemical consideration of the amount of substance. In chemistry a mole of one substance has different properties than a mole of another substance, while in thermodynamics a mole of any molecular substance produces the same pressure under the same temperature. Moles of different substances are exchangeable.

Usually mentioning the radian or other dimensionless units is supposed to help with the context of the quantity we are after, e.g., to remind that the frequency $\nu$ and the angular frequency $\omega=2\pi\nu$ are two related but not identical quantities. (`Cycles' the explicit use of which is recommended \cite{MP} is nothing else as a name of counted entities, which might be mentioned if helpful especially for a pure periodic motion.)

While considering the situation one has to remember that is not about cyclic motions, as often said. The question of a possibly new dimension and a new independent unit concerns a much broader range of phenomena. There are simple mathematical relations between exponential (with a pure imaginary argument) and trigonometric functions as well as between certain logarithmic and inverse trigonometric ones.
A function known on a finite interval can always be presented with the Fourier series, while a function on an infinite interval can be presented with a Fourier integral presentation. Whatever we do with trigonometric functions, by e.g., considering them as a function of a dimensional argument, immediately affects all the other functions through the Fourier transformation. It looks like any real and even certain cosmetic redefinitions should produce a great confusions in mathematical analysis and its application to basic physical and practical phenomena and may meet a strong opposition in public educational community. (See Appendix~\ref{s:list} for more details.)

Option \#1 turns actually to be transformed into two. Option \#1a is to consider the angle as a new base quantity and option \#1b is a terminological cloaking spell to cover option \#2 in such a way that it would look like option \#1 in certain applications.

As shown above we start both educationally and historically with a certain dual concept of the angle, which has been caused by a certain shortage of our understanding because both the educational and historical experience and knowledge come step by step. That is our choice whether we recognize that and allow those two concepts to merge as it is done in mathematics with the functional analysis or we introduce a chain of definitions to distinguish the concepts of the geometric angle and the functional one and to maintain their separation and parallel existence.

We remind that in contrast to the modification of the system of quantities as it was in the case of a transition from the Gaussian one to the SI, when the dimensions of various dimensional quantities have been changed, now we deal with a quantity, which is dimensionless in the functional concept and therefore will be still legitimate for the use within the SI even in case if the [geometric] angle will be treated as a dimensional one. On contrary, the old Gaussian quantities and units are not legitimate to be used within the SI.

A system, which allows for both concepts to be distinguished but be in a legitimate use, is a confusingly redundant one, because absolutely the same information can be delivered by using of the angle of either of the concepts without any [mathematical] ambiguity.

Mathematics is not a science, but a language of the science, and the geometry is a part of it. But it is a very special part. We study geometry of `our space' experimentally, which makes it a part of the physical world. However, `our' geometry and the `abstract' one cannot be separated. Once we make a plot of whatever we study we can measure details of the plot and in particular we can measure the angles with a protractor. In other words once we use a geometry as a language to discuss a problem, the angular quantities are involved. Any change in the treatment of the angle will immediately affect every area of physics and science where we make use of plots, which roughly means all the science.

E.g., once we consider complex number $a=4+3i$ we can make a plot (on a real piece of paper or directly on a real floor or a blackboard) and according to our knowledge from geometry $|a|=5$ and $\varphi_a=\pi/6$ is the phase of the complex number determined as an angle of a rectangular triangle with the legs of $3$ and $4$ arbitrary units. If we consider the impedance (at a certain frequency $\nu=\omega/2\pi$) with numerical values equal to $a$ in the SI units (i.e. $Z=a\,\Omega=(4+3i)\,\Omega$) then $|Z|=5\,\Omega$ and in the time domain the phase of the periodic current and voltage are different by $\pi/6$. We can consider a problem with the impedance as a problem with a consecutive connection of a resistor $R=4\,\Omega$ and an inductor with  $L\omega=3\,\Omega$. in such a case we can consider a plot in coordinates of $R-L\omega$ and make a conclusion on the phase difference in the time domain and ratio of the amplitudes of the voltage and current as $U_0/I_0=5\,\Omega$. We see that for certain problems the [geometric] angle in a `real' plane presenting the complex one, the phase of a complex number in the polar form, and the phase difference of two time series (for the current and voltage)
are exchangeable. The features related to the geometry may easily appear in a non-geometric problem. Since the `geometry' appears in non-geometric problems it does not look naturally to introduce an additional unit that cannot be expressed in terms of unity.

Personally, I do not think that we can afford to recognize the plane angle (and next the solid angle) as a new base quantity of the SI. We are also unlikely ready to introduce two parallel lines of the definitions of similar quantities and functions, the geometric ones and the functional ones, which eventually leaves us with option \#2, but probably in a combination with a certain version of option \#1b with a permissible non-SI unit for the angle.

Actually, the use of various grads and degrees is very similar to percents. The former are usually defined as submultiples (sometimes decimal (as for the Celsius temperature scale), sometimes nondecimals (as for the angles)) of a certain naturally existing reference value. When we speak about percents we remember to a certain extend that they are percents of something, however, when we speak about degrees and grads we rather do not mention that they are submultiple of some reference value. That gives to them a certain degree of independence and makes them dimensional by themselves, not as submultiples.

Perception of the angle as a dimensional quantity partly comes from our use of different units for it, such as the radian and the degree. The latter is defined as $1/360$th part of the full angle. Its dimension if any is determined by the dimension of the full angle that serves as the reference value. Roughly speaking, the full angle {\em is\/} the [natural] unit, while the degree is a submultiple of it. The reference to the full angle is a kind of omitted in any expressions and one may consider the degrees as a unit by itself. (In many applications ppm, promille, percents etc. are not considered as parts of a certain total (say, volume or population) but as the unit by themselves. The certain level of presence of alcohol in blood has legal consequences and the measurement device indicates it. What is the physical and chemical meaning of the units is unimportant for the involved sides.) The `true' dimension of the geometrical angle is unimportant for the customers, because to measure an angle in the degrees means to perform a measurement of its ratio to the full angle, but not of the absolute value of the angle. Making a decision that the full angle is a certain number ($2\pi$) does not affect any use of the degrees.

As concerning the radian that is again a question what it is and how we use it. The radian is the {\em name\/} of a special value of the central angle, such as the length of the related arc is equal to the radius. It is unimportant for the customers what is the meaning (if any) of such a special angle. Apart from the question of naming such an angle, the answer to the question what it is does not affect any measurements. If we decide that the value of the angle (the quantity value, not the numerical one) is just equal to unity, that would not affect the measurements and practice of using the radian for the angular measurements.

Any practical confusion comes from the practice of using the names of the angular units and have nothing to do with their geometrical, mathematical (in general), and physical meaning. Resolving possible confusions does not require a change of the SI system of quantities and units.

One may think that considering problems with the circular geometry or the periodic phenomena we may avoid confusions by introducing a dimensional angle. That is not that easy. There is a certain conflict in the perception of the angular and phase quantities that arises from two different views on them. One is focused on their `direct' measurement. When we measure an angle or a phase or the related quantities, such as their time derivatives, we prefer to recognize the related unit as a kind of independent ones, since our concern is mostly to compare one angle to another one.  (Actually, there are many areas where diverse submultiples of the full angle are used. The name of the submultiple is crucial to describe the `unit' of the measurements. Measuring a length in the millimeters, centimeters, and kilometers does not mean any departure from the SI. Even considering the foot, exactly equal to 0.3048\,m, allows for its use still remaining within the SI.)

The situation changes drastically when we study relations of the angular or phase-related quantities to the others, e.g., to mechanical linear quantities, be it for a rotation or a harmonic oscillation. Relating the acceleration and the radius (or displacement) is the equation of the second Newton's law, both sides of which (with the force and the kinematic acceleration) are free of the radians. However they determine the angular velocity (for the rotation) and angular frequency (for the periodic oscillation). I.e., the values of the angular quantities are naturally determined without any use of any independent angular units (see the Appendix~\ref{s:list} for details).

Metrological regulation serves to aim a number of objectives. One of them is to build an efficient coherent system of units and quantities, another is to facilitate for an efficient presentation of the results of measurements of various quantities. Customers interested in measurements of several individual quantities without their connections to the others and customers who are interested in quantities related to many others create two quite separate kinds of groups and they in general have different interest and different issues to address.

Possibly, the introduction of the dimensional angle would serve interests of the customers, involved in the angular and phase measurements, helping them to avoid possible confusions. Obviously, that would create numerous complications in the system of quantities and should require the redefinition of various mathematical and physical quantities, and changing a large number of fundamental relations. It would also end up with a system where a dimensional angle and its dimensionless twin coexist. The latter, being dimensionless, is a quantity that by definition is consistent with the SI system. There is no reason to expect that customers, that are not involved in angular and phase measurements, would accept the newcomer, such as the dimensional angle, that would make many equations more complicated.

\appendix

\section{Quantities related to the angle\label{s:phase}}

Discussing the angle one has to remember about a number of related quantities, such as the phase and the angular frequency.

1. In physics and practice we use two close terms, such as `the angle' and `the phase', which are exchangeable in many situations. When we consider a `real' rotation it is about a varying angle. However, when we describe the rotation the azimuthal angle in the polar coordinates plays a role of the phase in mathematical expressions, such as $x=\cos(\phi(t))$. As an example of an opposite situation, one can consider a complex number, say, $2+3i$ and plot it in the complex plane, allowing for its `Cartesian' and `polar' presentation. The phase of the complex number is an angle in the plane and can be directly measured by a protractor. Term `the angle' is the one that more closely relates to the geometric case in a `real' space, while `the phase' is often used in respect to the argument of periodic functions of the mathematical analysis, however, as we mention they are exchangeable in a sense. In this paper we prefer to use only term `the angle'.

2. When the phase changes with the time we introduce the angular frequency as its time derivative. When the angle does we introduce the angular velocity. The later may be understood as a scalar quantity equal to the time derivative of the angle in a plane, or as a vector (or rigorously speaking a pseudovector) the absolute value of which is the mentioned time derivative and the direction is along the rotation axis (for the counter clockwise rotation).

When we describe the rotation in the cartesian coordinates with expressions, such as $x(t)=\cos(\omega t)$, $\omega$ is the angular velocity (in its  scalar sense), but serves as the angular frequency for periodic change of $x(t)$. While the angular velocity and angular frequency are two different quantities they are exchangeable to a certain extend.

3. Fourier conjugated quantities appear through the Fourier transformation in numerous occasions in various areas of physics and practice. The transformation relates a pair of Fourier conjugated quantities through the phase of the involved integrands. Since they are related through the phase, once the phase is dimensional the dimensions of Fourier conjugated quantities are related through the unit of the phase or, since the angle and the phase are often exchangeable, through the unit of the angle.

(In simple words, frequency $\nu$ and time $t$ or angular frequency $\omega$ and $t$, depending of the definition of the Fourier transform, are Fourier conjugated quantities. Their dimensions are related through the condition of the dimensionless phase of the factors $e^{-i(2\pi\nu t)}$ or $e^{-i\omega t}$, depending on the definitions. The phase is dimensionless because once we do integrations or differentiations we have no other option as consider the exponential function as a dimensionless fucntion of a dimensionless argument.)

4. We often consider the geometry as exclusively geometry of our space. However, we love to make plots and to take advantage of geometric interpretations, such as considering the derivative as the tangential or the integral as the area below a curve. That automatically introduces angles to the problems that have nothing to do with geometry by themselves. However, such angles are real enough to be measurable with a protractor.

On the other hand, considering a function we can always have in mind that its argument is a kind of a time variable, which would allow for kinematic interpretation e.g. as the derivative as a velocity. The phase by itself is not related to the time evolution, but the frequency and the angular frequency are. Using kinematic interpretations in terms of possible cyclic motion (through the Fourier transformation) we rely on the phase.

If the functions and/or the defining equations are the same (the ones of interest, the ones for a geometric interpretation, and the ones for a kinematic interpretation) we can consider the interpretations as an analog models and make evaluations on those models that would involve angle- and phase-related quantities to diverse phenomena.  In `old good time' one could make a plot in a millimeter paper and cut out the area below the curve. Weighing it would give a result for the integral. In the meanwhile, weighing a piece of real paper means to use the descriptive geometry in our real physical world.

5. Speaking about geometry and calculation of various properties, we have to remember that diverse geometric factors appear in various identities. A geometric example are the ratio of the length of the arc and the radius, the ratio of the volume of the sector of the sphere and its radius etc. In physics we often have to sum or average certain values over all the directions. Such factors are dimensionless and they often are algebraic functions of the plane or solid angle. I.e., they present a natural set of dimensionless angle-related quantities, that have a clear geometrical or physical meaning, being directly measurable. Following the standard definition of the radian (as a name for unity) some of them are {\em equal\/} to the angles under the question. They are discussed in part in Sect.~\ref{s:a:geo}. On the other hand, the angle in degrees deals in fact the value of the portion of the full angle, which is also a dimensionless geometric factor.

\section{Basic identities with the phase and angle\label{s:list}}

Below we give a brief list of various identities mostly from the high school and basic University stage of the public education. They are related to diverse areas. The list is brief in a sense that we consider only several relations from each area, however our intention is to show the diversity of the fields involved. The relations are written using the angle measured in the radians assuming that the radian is just a name of unity and nothing else and therefore may be omitted.

The relations play a key role in a great number of various educational texts and reference books on physics and mathematics and their application to more practical issues. Because of various functional relations between trigonometric functions and others (through relations between trigonometric functions and exponential ones, transformations between Cartesian and polar/spherical/cylindrical coordinates, Fourier series and transformations, etc.) the identities cover a broad range of areas far from descriptive geometry and cyclic phenomena.

Each of those relations below can in principle be adjusted to a dimensional definition of the angle and to the angle measured in units other than the radian equal to unity. However, it would be very hard to adjust all of them in a coherent non-confusing way. We will need to edit the cornerstone relations that create the coherent system of quantities.

In contrast to a transition from the Gaussian system to the SI, the introduction of a dimensional angle would affect mathematical research community and, which may be even more important, mathematical educational community. The mentioned transition has changed units for a number of quantities, but as concerning the system of quantities the change was only for electrical phenomena. In the case of a consideration of quantity `angle' (and subsequently the `phase') that would affect all topical areas of physics in one or other way.

As we mentioned, most of these relations if not all can be in principle adjusted to the dimensional angle. However, there is another option that always exist. All the relations listed below are correct in a coherent sense. That means that apart of referring to the quantity under question as to `the angle', we can give it a different name and the relations will remain intact as a system of relations on quantity that `formerly' were referred to as the angle. The related quantity is dimensionless and therefore does not need any special approval to be used within the SI.

For references on the mathematical issues see \cite{arfken,menzel}, while on the physical ones see \cite{menzel} for more details.

\subsubsection{Geometry of a circle}

The length of an arc
\[
l=\theta r\,,
\]
and the related area of a sector in a circle
\[
s=\frac{\theta r^2}{2}
\]
are expressed through the same central angle $\theta$. So the angle can be determined from one of the identities (e.g., as $l/r$) and utilized to find the other. That makes the dimensionless ratio $l/r$ of practical importance, does not matter how we define the angle.

The angle in a triangle allows for a determination of trigonometric functions $\sin\theta$ and $\cos\theta$ etc. The introduction of the values of the functions does not require any specific parameterization of the angle, but some of their properties are.

Returning to the relations above we have to remember that they are important not only to compare the results of measurements of $\theta, r, l, s$, but also for many other relations. E.g., for the circular rotation we could intend to compare $d{\theta}/dt$ and $dl/dt$ (see below). Many relations in this appendix, being important by themselves, are of even more importance through consequences.

\subsubsection{Limits and derivatives}

One of the most famous limits studied at the high school is
\[
\lim_{x\to 0} \frac{\sin x}{x}=1\,.
\]
The limit plays a key role in derivation of the expressions for the derivatives of the trigonometric functions over their argument, the angle,
\begin{eqnarray}\label{eq:cos:prime}
\frac{d \sin x}{d x} = \cos x\;,\nonumber\\
\frac{d \cos x}{d x} = -\sin x\;.
\end{eqnarray}

With the sine and cosine as [geometric] functions of a dimension angle (of any dimension with the unit not equal to unity) the identities in (\ref{eq:cos:prime}) are incorrect, and even the dimensions in the LH and RH sides of the identities do not match. If we consider the frequency $\nu$ as measured in cycles per a unit of time, considering `cycles' as a kind of unit not equal to unity, then the argument of the sine and cosine should have the same dimension as $\nu t$, with the `unit' cycle. Such a function can be in principle introduces. However, the dimensions of the identities would not match as well. When the numeric functions go beyond simple algebraic ones, one has to be careful with the dimensions, otherwise many standard functional relations may be broken.

\subsubsection{Rotation and angular velocity\label{ss:rotation}}

The `most important circle' in the educational {\em physics\/} appears in the case of a circular periodic rotation around the center of the coordinates. It is described by vector equations
\begin{eqnarray}
\vec{v} &=& \vec{\omega}\times \vec{r}\;,\nonumber\\
\vec{a} &=& \vec{\omega}\times \vec{v}\;,\\
\end{eqnarray}
or by scalar ones
\begin{eqnarray}\label{eq:roration}
\omega&=&\frac{d\varphi}{dt}\;,\nonumber\\
v &=& \omega r\;,\nonumber\\
a &=& \omega v\;,
\end{eqnarray}
assuming that the acceleration in a point of trajectory is along the radius (in the opposite direction), while the velocity is perpendicular to the radius. Here $\vec{\omega}$ is the angular velocity and $\omega$ is its absolute value. The relation $v = \omega r$ immediately follows from one of the definitions of the angle as $l=\varphi r$ by its differentiation.

There is a difference in the perception of the description of the circular rotation depending on whether the focus is on a measurement of an angular quantity or on its relations to the linear ones. While measuring the angle or its time derivative, the angular velocity $\omega$, it seems convenient to introduce the units of the angle and angular velocity explicitly, and to treat them as independent to a certain extend. These units are not necessary to be related to the radian. However, while describing the circular rotation and considering various linear quantities ($r, v, a$), expressing one of them in the term of another and the angular velocity (as in (\ref{eq:roration})),
we are rather interested in the angular velocity in the radians per second with the radian equal to unity, because the relations between the mentioned linear kinematic quantities cannot include any units, but the meter and the second.

Successfully describing a circular trajectory, one may apply a similar notion to a description of a free motion of a point
in a rotating frame
\begin{equation}
\ddot{\vec{r}}=
- 2\left[\vec{\omega}\times\dot{\vec{r}}\right]
-\left[\vec{\omega}\times\left[\vec{\omega}\times\vec{r}\right]\right]\;.
\end{equation}
The equation gives us the Coriolis and centrifugal forces in the frame rotating with the angular velocity $\vec{\omega}$ in respect to the inertial frame.

The relations above are for the circular cyclic motion. However, the values related to the rotation, such as ${\omega}$ should not be understood as the ones related only to a circular cyclic motion. Once we have a plane motion, i.e., study a motion of Earth around Sun we may speak about the angular velocity. The standard Kepler's motion is a famous example of a non-circular rotation with a variable $\omega$. The planetary motion is {\em in vivo\/} not a circular one and because of a precession of the perihelion it is neither periodic. The influence of other planets on the `probe' one may produce an observable non-periodic component as well.

\subsubsection{Integration in polar, spherical, and cylindrical coordinates}

Use of polar, spherical, and cylindrical coordinates allows for a simplification in consideration of a rotation, however, it is also beneficial in case of certain symmetries in a problem of interest.

Here, we are interested not in a description of a motion in those coordinates, but in an integration over the volume
\begin{eqnarray}
\int {f(\vec{r})\, d^2\vec{r}}&=&  \int_0^\infty\int_0^{2\pi} {f(\vec{r})\, rdr d\varphi}\;,\nonumber\\
\int {f(\vec{r})\, d^3\vec{r}}&=&  \int_0^\infty\int_0^{2\pi}\int_{0}^{\pi} {f(\vec{r})\, r^2\sin\theta\, dr d\varphi d\theta}\;,\nonumber\\
\int {f(\vec{r})\, d^3\vec{r}}&=&  \int_0^\infty\int_0^{2\pi}\int_{-1}^{1} {f(\vec{r})\, r^2 dr d\varphi dt}\;,\nonumber\\
\int {f(\vec{r})\, d^3\vec{r}}&=&  \int_0^\infty\int_\Omega {f(\vec{r})\, r^2dr d\Omega}\;,\nonumber\\
\int {f(\vec{r})\, d^3\vec{r}}&=&  \int_0^\infty\int_0^{2\pi}\int_{-\infty}^{\infty} {f(\vec{r})\, rdr d\varphi dz}\;,\label{eq:int:3d}
\end{eqnarray}
where we use a standard notation and set $t\equiv \cos\theta$.

A specific case, closely related to the integration over the volume is an integration over the solid angle. The elementary solid angle is
\[
d\Omega = \sin\theta\, d\varphi d\theta = d\varphi d t=  \frac{d^3\vec{r}}{r^2dr}\,.
\]
Note, as a part of the integration over the volume in (\ref{eq:int:3d}), $d\Omega$ is dimensionless.

\subsubsection{Definite integrals}

There is a number of indefinite integrals which lead to $\arccos, \arcsin, \arctan$ (see, e.g, the one in (\ref{eq:arccos})). Their definite versions, such as
\[
\int_0^1{\frac{dx}{\sqrt{1-x^2}}}=\frac{\pi}2\;,
\]
lead to numerous $\pi$'s with rational coefficient. Many integrals, both definite and indefinite, can be evaluated numerically. They are just numbers. Calculating an integral numerically one cannot know whether it is related to inverse trigonometric functions and therefore is supposed to produce an angle (which may require the radians) or not. Note, the derivation of the integral above can be achieved easily by a substitution
\[
x=\sin\varphi
\]
with further integration over $\varphi$. While $\varphi$ is introduced as a pure mathematical trick one can explain the substitution geometrically as we would integrate in a plane.

Another integral which leads to $\pi$ is the one for the normalization of the Gaussian normal distribution
\[
\int_{-\infty}^{\infty} {\exp\left(-\frac{x^2}{2\sigma^2}\right)\,dx}=\frac{1}{\sqrt{2\pi}\sigma}\;.
\]
The appearance of $\pi$ is a geometric one in a sense. The famous derivation of the integral deals with
\[
\int_{-\infty}^{\infty} {\exp\left(-\frac{x^2}{2\sigma^2}\right)\,dx}\times \int_{-\infty}^{\infty} {\exp\left(-\frac{y^2}{2\sigma^2}\right)\,dy}
\]
being considered as an integral over the $x-y$ plane delivers $2\pi$ as the full angle while integrating in polar coordinates. Often a $2d$ distribution is plotted and considered `geometrically'. One may say that `building' a square of the integral and interpreting it as an integral over the plane is an `artificial trick' still unrelated to the geometry. We remind that the Maxwellian distribution on $\vec{v}$ has the form of the Gaussian one. The distribution on $v_x$ and $v_y$ is of the same form and one may consider an $2d$ distribution in the $v_x-v_y$ plane. That is an example where the same mathematical trick is related to a pretty real geometry.

Obtaining geometrical $\pi$'s we rather deal with the strait angle. Having $\pi$ as just a number in discussed above relations we consider the radian to be equal to unity.

For more integrals that involve $\pi$ see \cite{dwight}.

It is worth to mention that $\pi$ is also a part of the Euler formula
\begin{equation}\label{eq:Euler}
e^{-i\pi}=-1\;,
\end{equation}
which can be considered as an algebraic relation between three outstanding numbers ($\pi, e, i$) or pure geometrically in the complex plane.

\subsubsection{Harmonic oscillator\label{ss:harmonic}}

The harmonic oscillator is an important notion of classical physics that plays a great role in quantum mechanics and many other areas. The one with a spring is described by differential equation
\begin{equation}\label{eq:harm}
m\,\ddot{x}(t)=-k\,x(t)\;,
\end{equation}
That is a direct application of the 2nd Newton's law, where $x$ is a Cartesian coordinate, $m$ is the mass, and $-kx$ is the Hooke's-law force, which completely determines the dimension of $k$. Combining mentioned elements we can build a differential equation of a periodic motion (\ref{eq:harm}) and find
\begin{equation}\label{eq:harm1}
\omega^2=\frac{k}{m}\,,
\end{equation}
that expresses the angular frequency of the periodic motion through the {\em non-cyclic non-angular\/} quantities.

A similar situation is with many other oscillators when we derive their equations from somewhat general physical laws. In particular, equations for the electric oscillators are build of $R$, $L$, and $C$ which do not require any radians and $U, I, Q$ which involve them neither. Those quantities and relations between them are introduced for non-periodic effects and therefore they cannot involve any radians.

From the point of view of mathematical equations on well-defined physical quantities, equation
\begin{equation}\label{eq:harm:dis}
m\,\ddot{x}(t)=-\gamma \dot{x}(t)-k\,x(t)
\end{equation}
is the one for the damping oscillations. If one of the terms on the RH side is equal to zero it describes either the harmonic oscillations or the deceleration with a full stop due to friction in media. Some of the described motions are purely periodic, some are a kind of cyclic, but not periodic, some are not cyclic at all. The radian if presented should be presented in all three terms.

\subsubsection{Exponents and logarithms and the complex analysis}

The harmonic equation (\ref{eq:harm}) can be resolved in terms of imaginary exponents such as $\exp(\pm i\omega t)$ which is one of numerous examples of relations between exponents and trigonometric functions in mathematical analysis [of complex functions]. That is indeed not a high school level, but neither that is just University basics because some of the discussed here relations are among the most attractive in popular literature and spread in the underuniversity community.

We remind that the set of complex numbers $\mathbb{C}$ is not the real plane $\mathbb{R}^2$. The plane allows for fewer operations. Such an operation as a multiplication of two complex numbers $z_1z_2$ is not a standard one for $\mathbb{R}^2$. Nevertheless, it is very fruitful to rely on the plane. In particular, there is a complex analog of the polar coordinates
\[
z=re^{-i\varphi}\;.
\]
The phase $\varphi$ can be presented as an angle and measured by a protractor. To come to `Cartesian' presentation of a complex number
\[
z=x+i\,y
\]
we may use various semigeometric relations such as
\begin{eqnarray}\label{eq:e:cos}
e^{i\varphi}&=&\cos\varphi+i\sin\varphi\;,\nonumber\\
\cos\varphi&=&\frac{e^{i\varphi}+e^{-i\varphi}}2\;,\nonumber\\
\sin\varphi&=&\frac{e^{i\varphi}-e^{-i\varphi}}{2i}\;.
\end{eqnarray}

Geometry of a conus and trajectories of bodies in a central gravitation potential involve conic sections. For some of them it is beneficial to use hyperbolic trigonometric functions closely related to the trigonometric ones
\begin{eqnarray}
{\rm ch}(ix)&=&\cos x\,,\nonumber\\
{\rm sh}(ix)&=&i\,\sin x\,,\nonumber\\
\cos(ix)&=&{\rm ch}\,x\,,\nonumber\\
\sin(ix)&=&i\,{\rm sh}\,x\,.
\end{eqnarray}

The related identities for the inverse functions are
\begin{eqnarray}
\arcsin x&=&-i \ln\left(\sqrt{1-x^2}+ix\right)\,,\nonumber\\
\arccos x&=&-i \ln\left(x+i\sqrt{1-x^2}\right)\,,\nonumber\\
\arctan x&=&\frac1{2i}\ln\left(\frac{i-x}{i+x}\right)\,,
\end{eqnarray}
where we remind that all the functions ${\rm arc}..., \ln, \sqrt{...}$ above have multiple values and the branches on both sides of the identity should be specified. The logarithm-related functions on the RH side are those that are used as inverse hyperbolic trigonometric functions.

We see here a close relation between functions for cyclic phenomena with angles and phases and functions, that are not cyclic with real arguments but are cyclic with the pure imaginary ones and {\em vice versa\/}. In the functional analysis once a function is defined for pure real (or pure imaginary) argument the definition is extended to all the complex plane. Indeed, the involvement of the units cannot depend on whether the numerical value is real or imaginary.

\subsubsection{Natural line width, lifetime, resonances}

In a periodic motion we deal with $e^{-i\omega t}$.
If there is a line width or the emission involves the lifetime we have something like
\[
e^{-i(\omega-i\gamma/2) t}=e^{-i\omega t}\,e^{-(\gamma/2) t}\;,
\]
and therefore all the conventions on periodic motion and periodic functions, such as an exponent with the imaginary argument, are to be extended to a standard exponent of a real argument. That is related to any oscillating and wave phenomena (in mechanics, electricity, optics, acoustics, etc.) and quantum mechanics.

The requirement of classical physics for the appearance of such an exponent is the dissipation of energy (cf. (\ref{eq:harm:dis})), while in the case of quantum mechanics we speak about the decay and the lifetime. The exponential factor above combines together a cyclic parameter $\omega$ and a {\em non-cyclic\/} dissipation one $\gamma$.

\subsubsection{Fourier series and Fourier transformation}

In general, periodic functions are a powerful tool to study any dependence. E.g., if we have function $f(x)$ on the interval $[a,b]$ and $f(a)\neq f(b)$, we still can consider it as a periodic function with a period $2(b-a)$ and expand it into a Fourier series.

For a function defined on $(-\infty, +\infty)$ one can use the Fourier transformation. Two the most popular ones for the time series are
\begin{eqnarray}
\hat{f}(\nu)&=&\int {f(t)\, e^{-2\pi i \nu t}dt }\;,\label{eq:F:nu}\\
\tilde{f}(\omega)&=&\int {f(t)\, e^{-i \omega t}dt }\;.\label{eq:F:omega}
\end{eqnarray}
where the `frequency' $\nu$ and the `angular frequency' $\omega$ have dimension of $[t]^{-1}$. That follows not only from the exponentials but also from the inverse Fourier transformation which is an integration over the Fourier conjugated variable ($\nu$ or $\omega$) and has to deliver the initial function $f(t)$.

There is also the Laplace transformation
\begin{equation}\label{eq:laplace}
F(s)=\int_0^\infty {e^{-st}f(t)dt}\;,
\end{equation}
where ${\rm Re}(s)\geq0$ and in case ${\rm Re}(s)=0$ the Laplace transformation becomes a kind of the Fourier one. While the Fourier transform `speaks' in term of periodic functions the Laplace one does not.

\subsubsection{Taylor series}

Another very important series is the Taylor one. For the cosine it takes the form.
\begin{equation}\label{eq:cos:tay}
\cos x =\sum_{k=0}^\infty {(-1)^k\,\frac{x^{2k}}{(2k)!}}
%1-\frac{x^2}{2!}+\frac{x^4}{4!}=\dots
\;,
\end{equation}
while for the exponent the result is
\begin{equation}\label{eq:exp:tay}
e^x=\sum_{k=0}^\infty {\frac{x^{k}}{k!}}\;,
\end{equation}
which are closely related because of (\ref{eq:e:cos}). Taylor series for $\arccos x$ is given in (\ref{eq:arccos}).

Taylor series is one of the tool to expand the definition of the function. E.g., using the Taylor series of $\cos x$ and $\arccos x$ derived for the real numbers we can find those functions for pure imaginary ones.

The series can also be used to introduce functions of more complicated objects, such as operators. E.g., using the standard rules of the expansion of the exponent one can easily verify
\[
f(x+a)=\exp\left(a\frac{d}{dx}\right)\,f(x)\;,
\]
which is helpful for the introduction of the translation operator.

One of the key elements of quantum mechanics is the evolution operator (for a system)
\[
e^{-i{\cal H}t/\hbar}\;,
\]
where ${\cal H}$ is the Hamiltonian operator of the system.

Such operations play an important role at an advanced stage of the education. They are not studied at school or University basics, but it is important to understand, that the functions introduced at school are applied through all the education process and their properties should be consistent.

The Taylor series and/or functional relations can be applied also for a definition of the exponential of an Hermitian operators (${\cal A}$) to introduce operator $e^{i{\cal A}}$ which is a unitary operator. Such a relation between Hermitian and unitary operators plays an important role in many physical problems.

Different terms of the Taylor series of standard functions are dimensionless and therefore require dimensional arguments. Alternatively we may introduce dimensional derivatives and claim that the dimension of, say, $d \sin(x)/dx$ is not unity, which would be in contradiction with standard relations in (\ref{eq:cos:prime}).

\subsubsection{AC current and impedance}

We have already discussed the AC electricity in part (due to the harmonic oscillator---see Sect.~B\,f.).

The electric relations often involve time derivatives such as $I=\dot{Q}$. E.g., we can consider a capacitor with charge $Q=Q_0 \sin(\omega t + \phi)$, Since $I=\dot{Q}$ we find $I=I_0 \cos(\omega t + \phi)$ with $I_0 = \omega Q_0$, which does not leave a room for the radian as an independent unit.

From the dimensional point of view the complex impedance plays a role similar to the real resistance. Special cases are
\[
Z_R=R\,,~~Z_L=i\omega L\,,~~Z_C=\frac1{i\omega C}\,,
\]
where $C$ and $L$ are defined with {\em non-cyclic\/} phenomena in terms of the standard mechanical and electrical units, while the dimensions of all $Z_X$ is to be the same.\\

~\\

To conclude let's consider several advanced applications.

\subsubsection{Cauchy's residue theorem}

In functional analysis of the complex variables the contour of the integration can be deformed. In particular, the Fourier transformation starting as an integration over the real $t$ can go to complex values and the argument of the exponential factor instead of a pure imaginary one (which is associated with the phase) would receive a real part, which is not for a periodic function. In the complex functional analysis one often uses the theorem of residues.
E.g., the key relation between the Lorentzian profile in the frequency domain and the damping oscillations in the time domain is a result of such an integration. The profile and the damping oscillations are a part of consideration of mechanical, electrical and any other oscillators as well as of wave phenomena and quantum mechanics.

To relate the profile and the oscillations one has eventually to integrate around a pole in the complex plane. Similar is the integration over the fractionally rational functions. The integration around a pole delivers $2\pi$'s that appears as the {\em full angle\/} in the complex plane. That is a pure geometric source of many $\pi$'s while calculating definite integrals. The calculation of the integrals relies on their geometric sense clearly seen in a plot of the complex plane in our `real' world.

\subsubsection{Theoretical mechanics}

Rotations are also considered in a somewhat more sophisticated way than above in the theoretical [analytic] mechanics. An important quantity is the angular momentum of a system of interest
\[
\vec{M}=\sum_i\left[\vec{r}_i \times \vec{p}_i\right]\;,
\]
which is the standard definition of the angular momentum of the system in terms of the coordinates and momenta of its constituents.

Meanwhile, in the theoretical mechanics, considering it in terms of generalized coordinates and Lagrangian ${\cal L}$ we find
\[
M_z = \frac{\partial {\cal L}}{\partial \dot{\varphi}}\;,
\]
where $\varphi$ is the polar angle in the $xy$ plane for the positioning of the center of mass of the system. In the meantime, we also write $\omega_z=\dot{\varphi}$.

The analytic mechanics in the Lagrangian form works mostly with coordinates, like $\varphi$, and their velocities like $\dot{\varphi}$, while the Hamiltonian mechanics expressed their evaluations in terms of coordinates and their momenta like $M_z$. All three types of quantities are closely related as well as their dimensions and units are.

\subsubsection{Generalized Euclidean space}

Linear algebra of an Euclidean space (= a vector [linear] space + a scalar product) is often considered as a non-true geometry, in contrast to the `real' one.

The vector space $\mathbb{R}^n$ with a scalar product satisfies conditions that each scalar product $(A\cdot B)$ is a real number, the value of $(A\cdot A)$ is a positive one (except of a special case of zero vector, when it is equal to zero) and
\[
\left\vert\frac{(A\cdot B)}{\sqrt{(A\cdot A)}\,\sqrt{(B\cdot B)}}\right\vert \leq 1\;,
\]
which allows for an introduction of a certain angle $\theta$ such as
\[
\frac{(A\cdot B)}{\sqrt{(A\cdot A)}\,\sqrt{(B\cdot B)}}=\cos\theta\;.
\]

If $n=2,3$ we can `realize' such an abstract space in `our' space and do direct geometric measurements. If $n\geq4$ we can still make projections into our space or in a plane in our space or study related cross sections. As we discussed for the complex numbers, the angles are `real' angles because we can draw them (with projections if necessary) and measure. We may also build an analog model with, e.g., several entangled oscillators, which have total dof$\;\geq 4$ and measure the phase. In general, any abstract construction which is used in physics is supposed to allow for measurements.
%`Abstract angles' usually allow for measurements as certain `real' angles or `real' phases.

The Euclidean space is an advanced notion while working algebraically, that is why we like to use our geometric perception both for education and for practical evaluations. Speaking about `real' and `imagination' space, let's remember, that we live in an $3d$ space but love to use various $2d$ projections, because our geometric intuition often better works when we see a plot on a piece of paper or a screen. We widely use $2d$ projections of various $3d$ problems. Now with computers we may easily plot a line in a $3d$ case. In the earlier time we would plots the lines in the $2d$ projections instead. The `reality' of the geometric space is a tricky issue. As we mentioned above we can measure the phase of a complex number by a `real' protractor once we use a plot of a complex plane in a piece of paper in our `real' world or just on a floor.

\subsubsection{Base quantum mechanics}

Base quantum mechanics is not only basics of contemporary physics but also a topic of many popular projects.

%Evolution operator

The quasiclassical approximation deals with $e^{iS/\hbar}$. This factor plays a crucial role in proving that the classical mechanics is a certain limit of the quantum one.

The Feynman integral over the trajectories also relies on this factor \cite{feynman}. That is one of several equivalent formulations of quantum mechanics (along with the matrix mechanics by W. Heisenberg and the wave one by E. Sch\"odinger).

Quantum mechanics often deals with
\[
\exp\left(-i\frac{E\cdot t}{\hbar}\right)
\]
and
\[
\exp\left(+i\frac{\bigl(\vec{p}\cdot\vec{r}\bigr)}{\hbar}\right)\;.
\]
Those factor are used for the introduction of the operators
\begin{eqnarray}
{\cal H}=i{\hbar}\frac{\partial }{\partial t}\,,\nonumber\\
\hat{\vec{p}}=-i{\hbar}\frac{\partial }{\partial  \vec{r}}\;.
\end{eqnarray}
The dimensions of the operators have been already defined through the classical limit.

It is worth to mention that in %advanced
quantum mechanics
\[
\hbar=\frac{h}{2\pi}
\]
was introduced because of numerous appearances of this combination in important identities such as
\[
\Delta p \, \Delta x \leq \hbar/2
\]
or
\[
L_z = m_z \hbar\;,
\]
where $L_z$ is the projection of the orbital momentum and $m_z$ is the azimuthal quantum number.
In some of those expressions a discussion on the radian as a part of the unit for $\hbar$ cannot even appear. The simplifications with $\hbar$ takes place because of the existence of two Fourier transformations (\ref{eq:F:nu}) and (\ref{eq:F:omega}). As long as we concern about the frequency of a periodic motion, $\nu$, and related to it $h=E/\nu$ seem the most natural, however if we have to deal with Fourier transforms the natural quantity is rather $\omega$ and related to it $\hbar=E/\omega$.
As discussed above the issue on the angle and phase because of the Fourier transformation starting with the pure circular and/or periodic phenomena propagates through all the physics.

%\subsubsection{}\subsubsection{}\subsubsection{}

\subsubsection{Special functions and differential equations}

Special functions play a key role in mathematical description of various physical problem. They are often introduced as solutions of certain differential equations. One can consider the trigonometric functions as solutions of the problem of the harmonic oscillator (cf. (\ref{eq:harm})). A big variety of special functions can be presented in terms of hypergeometric function ${}_2F_1(a,b,c;z)$ and confluent hypergeometric one  ${}_1F_1(a,b,z)$ (see, e.g., \cite{whittaker,abramowitz}). That sets one more group of relations between exponents, trigonometric functions, polynomials etc. since the same `generic' function describes periodic and diverse aperiodic phenomena.

\subsubsection{Quantity equations and equations with quantity values}

In order to avoid the problem with the use of a dimensional angle $\theta_d$ and a dimensionless one, the later is sometimes considered as its numerical value $\{\theta_d\}$ in the radians, but not as an independent quantity. That looks like a solution to avoid a redundance with a simultaneous appearance of $\theta_d$ and its dimensionless twin.

However, it creates a problem in formal metrology. That is crucial for structure of metrological relations, that we consider two kinds of the relations (see, e.g., \cite{VIM}).
\begin{itemize}
\item One is on quantities. Such relations do not depend on units. They contain quantities themselves, but neither their unit-dependent numerical values nor the units.
\item The other group is on numerical values of quantities. They do depend on the choice of units.
\end{itemize}
The only way for a numerical value to appear in a quantity equation is the case of a dimensionless quantity when the quantity is identical to its numerical value while the unit is a special name for unity and nothing else. Saying `nothing else' we mean the mathematical meaning of the unit as an entry of an equation. Use of traditional special names for unity plays in this sense rather a social role as a tool to avoid confusions, but is not required mathematically.

Technically that means that we cannot include $\{\theta_d\}$ into a quantity equation\footnote{One of the examples of a possible mess is the time derivative $d\{\theta_d\}/dt$, which may be required to make the argument of $\sin(\omega t)$ dimensionless. We have to introduce a hybrid, which combines a numerical value and a [dimensional] quantity.}, such as many equations described above in the appendix.  However, we may include the dimensionless angle as a separate well-defined quantity into it.

That would create a problem. The mechanical equations of motion in Cartesian coordinates neither include specific angular units nor cycles. However, they are sufficient to completely describe harmonic oscillator (see, e.g., (\ref{eq:harm})) and therefore its angular frequency and frequency. In part that is the same situation as in descriptive geometry. We are capable to completely describe the angle by values of the dimensionless sine and cosine.

There is another important issue which is often missed. The question about the angle is not a question about advantages and disadvantages of its treatment as a dimensional quantity. The dimensionless angle, presented in the expressions above {\em does\/} already exist. It is a legitimate quantity in mathematical sense and as a dimensionless one it {\em cannot\/} contradict to the SI. E.g., one can determine the value $l/r$ measuring the length of an arc and the radius (in the SI units) and express through it the area of the related sector (also in the SI units) as
\[
S=\frac12 \frac{l}{r} r^2\;.
\]
A similar situation, as we just reminded is in mechanics. The mechanical equation of the harmonic oscillator (\ref{eq:harm}), that completely determined the angular frequency allows one to express it, but it can do that only in terms of units of the length, time, and mass (cf. (\ref{eq:harm1})), because no independent specific angular units are not present in the equation.
%Forces and consequently accelerations in Newtonian mechanics codescribes in Cartesian coordinates all periodic and
%circular motions do involves any radians.

So the dimensionless angle, determined in the geometric example as $l/r$, is not only a legitimate SI quantity but also a useful one that appears naturally. The very use of the dimensionless angle, {\em consistent\/} with the SI, cannot be forbidden by a metrological regulation. Technically that means that we are to introduce an additional dimensional quantity with a related unit, assign the name `angle' to it, and {\em recommend\/} to use it in certain situations, while the dimensionless angle remains to exist. We may assign a different name to the latter, but we cannot really recommend to stop its use, which would rather lead to more confusions than to less.

It is often said that the use of the degrees and the radians is traditional. Yes. It is. But the use of a dimensionless angle and the angle measured in terms of the radian, which is equal to unity, is traditional as well. It is studied at high school and presented in pretty old mathematical books. While the tradition of the use of various units for the angle is older than the use of a dimensionless angle, the later has already successfully embedded in a much more broad area of scientific applications.

The common practice is the use of `radian' when necessary to avoid confusions. We usually say that an angle is equal to 2 radians, not to 2. But we also say that an angle is $\pi/4$, knowing that $\pi$ is just a number, but it is usually used for the angle in the radians and not with degrees. We also often use that in equations. Such a mixed use of the radian indicates that in general we do understand that it is a name for unity starting from the high-school education level.

%\newpage

\section{Would-be dimensional angle and dimensionless geometric parameters\label{s:a:geo}}

As shown in the previous section of the appendix, many expressions for both the base law of Nature and the equations of measurements contain various geometric factors. For a large number of quantities they are dimensionless and measurable. If we make the angle dimensional that would technically mean that in identities, that do not contain explicitly any angular variables, such as many of considered above, we have to introduce a dimensionless geometric parameter. That may be done with help of the full angle. For the plane angle $\theta$ with the full [plane] angle denoted as $\theta_{\rm full}$ such a dimensionless geometric parameter would be
\[
\theta'=\frac{2\pi}{\theta_{\rm full}}\,\theta\;.
\]
This parameter would enter series for the trigonometric functions, various integrals and derivatives, Fourier transformations, phase factors of the complex numbers, relations between the angular and linear velocities etc. It is completely compatible with the angle measured in the radians with the radian being a name of unity.

The origin of such a parameter is obvious for those who knows of the dimensionless angle and that the dimensionless angle {\em naturally\/} appear in many equations of physics and mathematics. While many who understand the origin of such a quantity may wonder why we need it, those who does not understand may wonder why we often have factor of $2\pi/\theta_{\rm full}$.

The values of the full angle $\theta_{\rm full}$ should be treated as new dimensional fundamental constant and can be measured in any units of our choice. That would create a self-consistent and mathematically valid system of units and quantities.

Apparently, if we choose the angle to be dimensional such a dimensionless combination and some of its derivatives, that are measurable and have a direct physical and/or geometric meaning, would deserve special names. They would be in a massive use since they allows for a simplification of many equations in many areas. Simultaneous use of $\theta$ and $\theta'$ would likely produce many confusions.

The standard approach considers a vector (including the position vector) as an object that is determined by its magnitude and its direction. We often need to deal rather with the directions than with the angles, e.g., to integrate over them. The most straightforward way to introduce the integration over the directions [of the position vector] in the $3d$ space is to deal with the angular integration as
\[
d\Omega_{\rm direction}=\frac{d^3r}{r^2dr}\;.
\]
The integration over the directions is dimensionless. Introducing the dimensional angle we open in a sense the question whether the integration over the directions is the integration over the angles or not, which relates to both plane and solid angles.
That would also create a room of confusions.

%\newpage

\onecolumngrid

\end{document}